\documentclass[pdflatex,sn-nature]{sn-jnl}

\usepackage{graphicx}
\usepackage{amsmath,amssymb,amsfonts}
\usepackage{booktabs}
\usepackage{xcolor}
\usepackage{rotating}
\usepackage{float}
\usepackage{fancyvrb}

\DefineVerbatimEnvironment{PromptBlock}{Verbatim}{
  fontsize=\footnotesize
}

\begin{document}


\title[AI-Assisted Goal Setting]{AI-Assisted Goal Setting Improves Goal Progress Through Social Accountability}

\author*[1]{\fnm{Michel} \sur{Schimpf}}\email{ms2957@cam.ac.uk}

\author[2]{\fnm{Julian} \sur{Voigt}}

\author[1]{\fnm{Thomas} \sur{Bohn\'{e}}}

\affil*[1]{\orgdiv{Department of Engineering}, \orgname{University of Cambridge}, \orgaddress{\city{Cambridge}, \country{United Kingdom}}}

\affil[2]{\orgname{Technical University of Munich}, \orgaddress{\city{Munich}, \country{Germany}}}


\abstract{%
Helping people identify and pursue personally meaningful career goals at scale remains a key challenge in applied psychology. Career coaching can improve goal quality and attainment, but its cost and limited availability restrict access. Large language model (LLM)-based chatbots offer a scalable alternative, yet the psychological mechanisms by which they might support goal pursuit remain untested. Here we report a preregistered three-arm randomised controlled trial ($N = 517$) comparing an AI career coach (``Leon,'' powered by Claude Sonnet), a matched structured written questionnaire covering closely matched reflective topics, and a no-support control on goal progress at a two-week follow-up. The AI chatbot produced significantly higher goal progress than the control ($d = 0.33$, $p = .016$). Compared with the written-reflection condition, the AI did not significantly improve overall goal progress, but it increased perceived social accountability. In the preregistered mediation model, perceived accountability mediated the AI-over-questionnaire effect on goal progress (indirect effect $= 0.15$, 95\% CI $[0.04, 0.31]$), whereas self-concordance did not. These findings suggest that AI-assisted goal setting can improve short-term goal progress, and that its clearest added value over structured self-reflection lies in increasing felt accountability.
}


\maketitle

\noindent\textbf{Preprint.} This manuscript has not undergone peer review.

\noindent\textbf{Version:} March 17, 2026.

\medskip


\section{Introduction}\label{sec:intro}

Structured support for setting and pursuing goals improves performance and goal attainment. Goal-setting theory argues that clear goals help direct attention, mobilise effort, sustain persistence, and prompt effective task strategies~\cite{locke1990theory,lockelatham2002}. A distinct and complementary driver of goal attainment is \emph{social accountability}: making goals visible to others---through public commitment, progress sharing, or the oversight of a coach---reliably amplifies attainment. A meta-analysis of 138 independent experiments found $d = 0.40$ for progress-monitoring interventions overall, with substantially stronger effects when monitoring was combined with social commitment or public reporting~\cite{harkin2016does}. Yet structured support for this process remains inaccessible for most working adults.

Career coaching is among the most effective structured interventions for professional development and goal pursuit, with a meta-analysis of coaching RCTs finding a mean effect of Hedges' $g = 0.59$ across performance, well-being, and goal-attainment outcomes~\cite{dehaan2023}. A skilled coach helps clients articulate meaningful goals while simultaneously creating the relational accountability that sustains commitment~\cite{terblanche2022comparing}. However, professional coaching is expensive and access is highly stratified---concentrated among managers and executives and largely unavailable to the majority of working adults who would benefit. Large language model (LLM)-based chatbots have emerged as a technically scalable candidate for delivering structured goal-setting support at population scale. Unlike earlier rule-based systems, modern LLMs can conduct open-ended, contextually sensitive, and personalised conversations at negligible marginal cost---with evidence that they can match or surpass human interlocutors in persuasive impact~\cite{salvi2025} and show effectiveness across psychological intervention domains including therapy, health behaviour change, and educational coaching~\cite{demszky2023using,fitzpatrick2017delivering,aggarwal2023artificial,kuhail2023interacting}.

In goal-setting applications specifically, AI coaching can improve goal attainment over extended periods~\cite{terblanche2022efficacy,terblanche2022comparing}. However, no study has formally isolated the \emph{psychological mechanisms} through which AI-assisted goal setting produces downstream progress---and no preregistered randomised trial has tested the effect of a modern LLM-based chatbot on goal progress at follow-up in working adults.

Two preregistered mechanisms offer theoretically distinct and independently testable accounts of how AI-assisted goal setting might influence downstream goal progress. First, \emph{social accountability}: the felt obligation to justify one's choices and actions to a perceived evaluator is a well-established driver of goal follow-through~\cite{frink1998toward}. A comprehensive review of the accountability literature identifies awareness of an evaluative audience as the primary antecedent, operating through increased vigilance and public commitment~\cite{lerner1999accounting}; the meta-analytic evidence confirms that these processes reliably increase goal attainment~\cite{harkin2016does}. Critically, the evaluative audience need not be human: the Computers Are Social Actors (CASA) paradigm demonstrates that people spontaneously apply social norms and expectations to machines that communicate in natural language, treating them as social entities worthy of reciprocation~\cite{nass1994computers,nass2000machines}. A chatbot that conducts a personalised goal-setting dialogue, expresses interest in future progress, and schedules an explicit follow-up check-in may therefore generate the felt accountability that has previously required a human partner~\cite{mohr2011supportive}. Second, \emph{self-concordance}: self-determination theory proposes that goals pursued for autonomous, intrinsic reasons produce greater sustained effort~\cite{deciryan2000what,sheldon1995self,sheldon1999role}; a guided conversation surfacing personal values and priorities may shift goal motivation toward more autonomous regulation.

We tested these questions in a preregistered three-arm randomised controlled trial ($N = 517$ randomised). Participants were assigned to: (G3) an AI career coach powered by Claude Sonnet that guided a structured goal-setting conversation; (G2) a matched written questionnaire covering the same broad reflective topics---career background, energising activities, priorities, and goal drafting---in a static, non-interactive format; or (G1) a no-support control. The questionnaire was designed as an active control that matched the AI condition on broad reflective topics while differing in conversational format. The AI-versus-questionnaire contrast therefore tests the added value of conversational interaction relative to a static reflection exercise, while not constituting a perfectly single-ingredient manipulation. Goal progress was assessed at a two-week follow-up. Preregistered hypotheses tested main effects on goal progress (H1a--c), effects on accountability and self-concordance as mediators (H2a--b, H3), and mediated pathways (H4).


\section{Experimental Design}\label{sec:design}

$N = 517$ employed adults were recruited via Prolific and randomly assigned to one of three conditions in a preregistered between-subjects randomised controlled trial (preregistration: \url{https://archive.org/details/osf-registrations-6awmq-v1}). Eligibility criteria were UK or US residence, age 18--50, current employment (full- or part-time), and English fluency. The study was delivered via a custom mobile application across two online sessions approximately 14 days apart; the participant flow is shown in Appendix Fig.~\ref{fig:consort}.

\begin{figure}[H]
  \centering
  \includegraphics[width=0.85\textwidth]{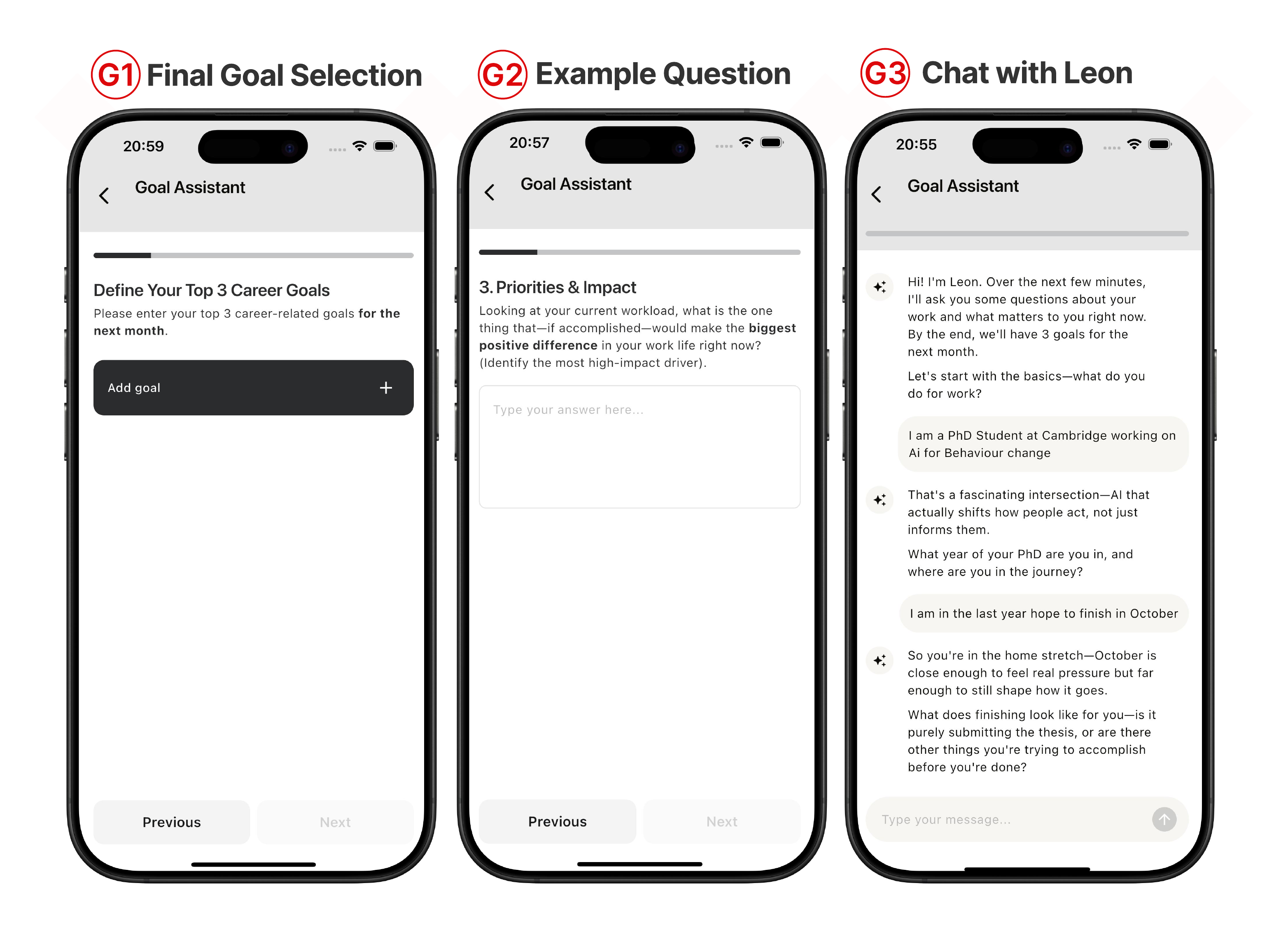}
  \caption{Screenshots of the study application across the three conditions. \textbf{G1}~(left): the final goal-entry screen, where participants typed their goals directly. \textbf{G2}~(centre): an example structured reflection prompt (``Priorities \& Impact''), one of five written questions completed before goal entry. \textbf{G3}~(right): a mid-conversation exchange with ``Leon,'' the AI career coach, probing the participant's work situation before collaborative goal formulation.}
  \label{fig:app}
\end{figure}

At Time~1 (T1), participants completed two background measures (AI knowledge; generative AI use frequency) and were instructed to set three career-related goals for the coming month via their assigned condition:
\begin{itemize}
  \item \textbf{G1~(Control):} Participants entered their goals directly with no reflective support.
  \item \textbf{G2~(Questionnaire):} Participants completed five structured open-ended written prompts---career background, energising activities, key priorities, upcoming constraints, and a goal drafting exercise with a should-versus-want check---before entering their goals. A 20-minute minimum engagement time was platform-enforced.
  \item \textbf{G3~(AI Chatbot):} Participants engaged in a guided conversation with ``Leon,'' an AI career coach powered by Claude Sonnet, which led them through four structured phases (background exploration, energising activities, priorities, and collaborative goal formulation). Sessions lasted $M = 21.9$~min ($SD = 9.7$; $M = 48.9$ messages).
\end{itemize}
Representative screenshots of the G1 (Control) and G3 (AI Chatbot) interfaces are shown in Fig.~\ref{fig:app}. The questionnaire and AI conditions were deliberately matched on broad reflective topics while differing in conversational format.

 Immediately after goal entry, all participants completed T1 measures: accountability, self-concordance (per goal), goal commitment, manipulation checks (perceived interactivity; perceived structured reflection), and Net Promoter Score (NPS).

At Time~2 (T2; actual gap: $M = 14.4$ days, $SD = 1.5$), participants were reminded of their T1 goals and completed the primary outcome---goal progress, assessed with three items per goal (9 items; $\alpha = .86$; 7-point scale)---along with T2 self-concordance, accountability, enthusiasm, goal commitment, and NPS. Participants who passed both embedded attention checks at T1 and returned for T2 comprised the analytic sample ($n = 323$; G1: $n = 105$; G2: $n = 111$; G3: $n = 107$).

The two preregistered mediators were accountability and self-concordance measured at T1. Goal specificity was assessed post hoc via LLM-based coding blind to condition. Main effects on goal progress (H1a--c) were tested with planned pairwise Welch's $t$-tests; mediation (H4a--c) with a parallel mediation framework using 5,000 bootstraps~\cite{hayes2013introduction}. Full participant sampling criteria, measures, and statistical specifications are provided in the Methods section.


\section{Results}\label{sec:results}

\paragraph{Sample.}
$N = 517$ participants were randomised to condition; 323 were included in the analytic sample (G1 Control: $n = 105$; G2 Questionnaire: $n = 111$; G3 AI Chatbot: $n = 107$), corresponding to 62.5\% of all randomised participants and 75.5\% retention from T1 to T2 among the 428 participants who passed the T1 attention check (Appendix Fig.~\ref{fig:consort}). Dropout from T1 to T2 was not differential across conditions ($\chi^2(2) = 0.70$, $p = .704$), and completers did not differ from non-completers on AI knowledge ($p = .097$) or generative AI use ($p = .182$). Randomisation was successful: conditions did not differ on age, gender, AI knowledge, or generative AI use frequency (all $p > .26$). The sample was predominantly female (63.2\%), with a mean age of 34.7 years ($SD = 7.9$), recruited from the United Kingdom (61.6\%) and United States (38.1\%), and primarily in full-time employment (74.6\%). All primary analyses were preregistered at \url{https://archive.org/details/osf-registrations-6awmq-v1}. Descriptive statistics and intercorrelations are in Table~\ref{tab:desc}.

\begin{table}[h]
\caption{Descriptive statistics and intercorrelations for primary study variables ($N = 323$; goal specificity $n = 292$).}\label{tab:desc}
\begin{tabular}{lrrrrrrr}
\toprule
Variable & $M$ & $SD$ & 1 & 2 & 3 & 4 & 5 \\
\midrule
1. Goal Progress (T2)    & 3.27 & 1.33 & — & & & & \\
2. Accountability (T1)   & 3.51 & 1.27 & $.19^*$ & — & & & \\
3. Self-Concordance (T1) & 1.91 & 2.45 & $-.07$ & $-.01$ & — & & \\
4. NPS (T1)              & 5.68 & 2.65 & $.30^*$ & $.50^*$ & $.17^*$ & — & \\
5. Goal Specificity      & 1.26 & 0.96 & $.14^*$ & $.09$ & $-.04$ & $.18^*$ & — \\
\bottomrule
\end{tabular}
\vspace{2pt}
{\footnotesize $^*p < .05$. Goal Progress 1--7; Accountability 1--7; Self-Concordance $-8$ to $+8$; NPS 0--10; Goal Specificity 0--4 (LLM-coded).}
\end{table}

\paragraph{Manipulation checks.}
The AI chatbot was rated substantially more interactive than both the questionnaire and control ($F(2, 320) = 157.10$, $p < .001$; G3: $M = 5.07$; G2: $M = 1.97$; G1: $M = 1.60$). The AI and questionnaire conditions were perceived as equally involving structured reflection, both higher than the control ($F(2, 320) = 86.78$, $p < .001$; G3: $M = 4.45$; G2: $M = 4.13$; G1: $M = 1.85$). This supports the intended design: the AI and questionnaire conditions differed strongly in perceived interactivity while being similar in perceived structured reflection.

\begin{figure}[h]
  \centering
  \includegraphics[width=0.96\textwidth]{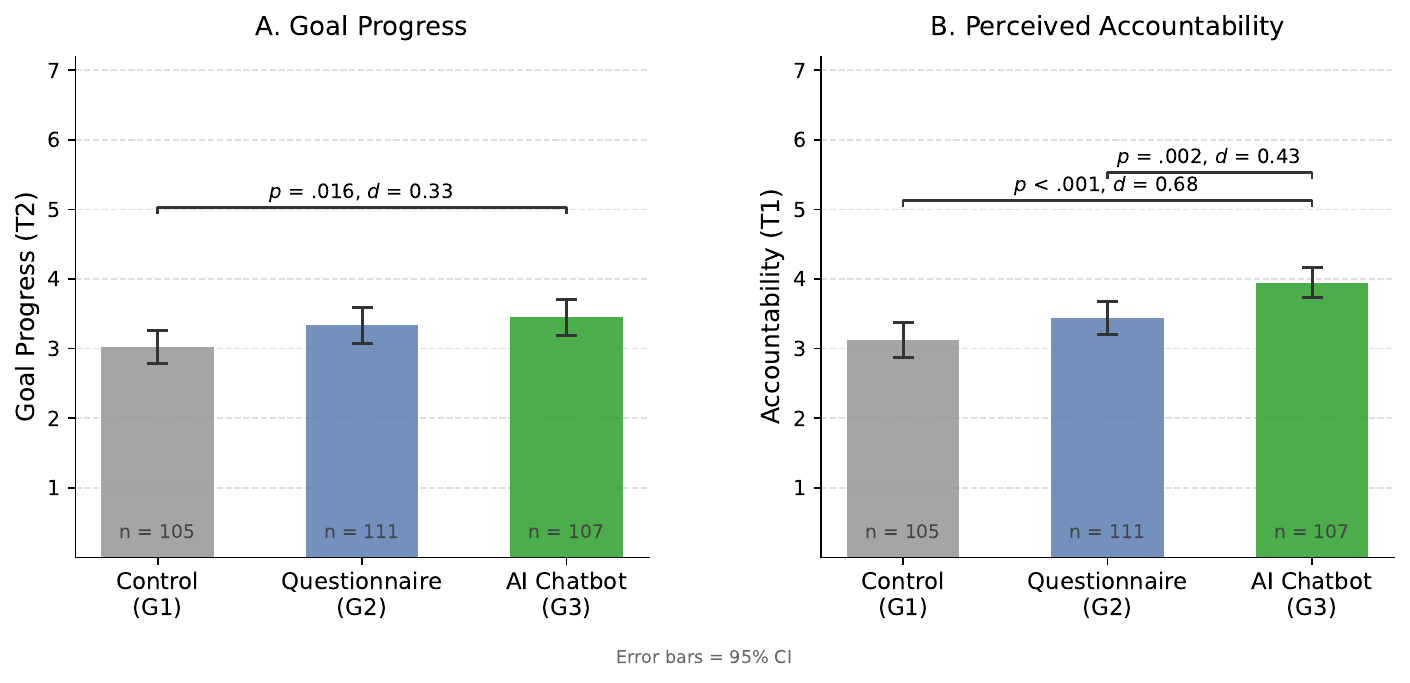}
  \caption{Primary outcome and main mediator by condition. Panel~A shows goal progress at Time~2; Panel~B shows perceived accountability at Time~1. Bars show condition means; error bars are 95\% confidence intervals. Key preregistered contrasts are annotated. $n$ per condition is shown within each panel.}
  \label{fig:progress}
\end{figure}

\paragraph{AI coaching improved goal progress.}
The AI chatbot produced significantly higher goal progress at two weeks than the control ($M_{\text{AI}} = 3.45$, $SD = 1.35$ vs.\ $M_{\text{Control}} = 3.02$, $SD = 1.22$; $t = 2.42$, $p = .016$, $d = 0.33$; Fig.~\ref{fig:progress}, Panel~A). The questionnaire showed a non-significant trend relative to the control ($M_{\text{Quest}} = 3.34$; $d = 0.24$, $p = .076$), and the AI and questionnaire did not differ significantly ($d = 0.08$, $p = .540$). A preregistered ANCOVA with age, gender, AI knowledge, and generative AI use frequency as covariates confirmed the AI-versus-control advantage ($b = 0.40$, $p = .032$, 95\% CI $[0.03, 0.76]$; $N = 317$; six participants were excluded due to missing age in the Prolific demographic export, including one whose record was consent-revoked post-completion). Cook's distance analysis ($4/n$ threshold) flagged 22 cases (max $D = 0.054$); inspection revealed no implausible values and no observations were removed.

\paragraph{Accountability.}
The AI chatbot generated substantially more perceived accountability than both the questionnaire ($d = 0.43$, $p = .002$) and the control ($d = 0.68$, $p < .001$; $F(2, 320) = 12.16$, $p < .001$, $\eta^2 = .071$; Fig.~\ref{fig:progress}, Panel~B). The questionnaire did not differ significantly from the control ($d = 0.24$, $p = .074$).

\paragraph{Self-concordance.}
Overall self-concordance did not differ across conditions ($F(2, 320) = 0.34$, $p = .709$; all pairwise $d \leq 0.12$). An exploratory decomposition by goal domain within G3 suggested heterogeneity: participants who surfaced at least one non-career goal ($n = 70$) showed higher self-concordance than those who set only career goals ($n = 37$) at both T1 ($d = 0.44$, $p = .033$) and T2 ($d = 0.53$, $p = .011$; Appendix Table~\ref{tab:sc_domain}).

\paragraph{Mediator--outcome associations.}
In the preregistered H3 model regressing goal progress on T1 accountability, T1 self-concordance, and condition dummies, accountability predicted higher goal progress ($b = 0.17$, $SE = 0.06$, $p = .004$), whereas self-concordance did not ($b = -0.04$, $p = .182$; $R^2 = .049$, $n = 323$).

\begin{figure}[h]
  \centering
  \includegraphics[width=0.82\textwidth]{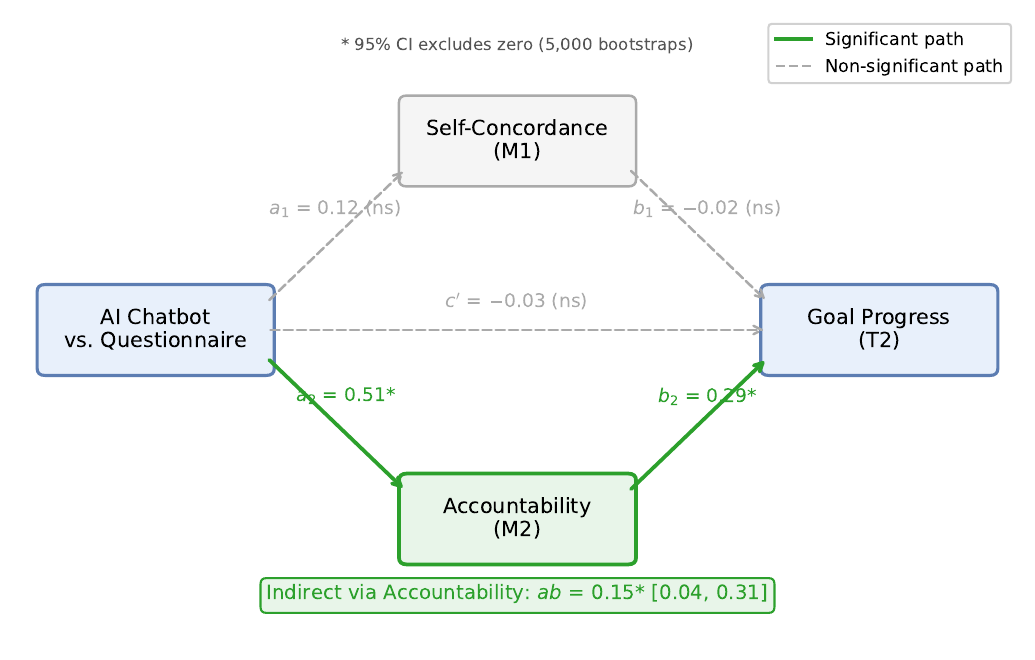}
  \caption{Preregistered two-mediator parallel mediation model for the AI Chatbot versus Questionnaire contrast. Green paths are statistically significant (95\% CI excludes zero, 5,000 bootstraps); grey dashed paths are non-significant. Path coefficients shown are unstandardised. The indirect effect through Accountability ($ab = 0.15$, 95\% CI $[0.04, 0.31]$) indicates mediation of the AI-versus-Questionnaire contrast; the indirect effect through Self-Concordance was negligible ($ab = -0.002$, 95\% CI $[-0.06, 0.04]$, ns).}
  \label{fig:mediation}
\end{figure}

\paragraph{Mediation.}
Preregistered two-mediator mediation analyses yielded different patterns across the three planned contrasts. For AI versus Control (H4a), the total effect was significant ($c = 0.43$, $p = .016$), but neither the accountability nor self-concordance indirect effect reached significance and the direct effect remained ($c' = 0.39$). For Questionnaire versus Control (H4b), neither indirect path nor the total indirect effect was significant. For AI versus Questionnaire (H4c), accountability mediated the contrast ($ab = 0.15$, 95\% CI $[0.04, 0.31]$), whereas self-concordance did not; the direct effect was near zero and non-significant ($c' = -0.03$; Fig.~\ref{fig:mediation}).

\subsection*{Exploratory analyses}
The following analyses were not preregistered and should be interpreted as exploratory.

\paragraph{Goal specificity.}
Goal specificity was assessed post hoc via LLM-assisted coding of time-frame and measurability dimensions~\cite{schimpf2025goal}. In a clustered goal-level model (one row per goal, standard errors clustered by participant), AI chatbot goals were substantially more specific than goals in both the control ($b = 1.29$, 95\% CI $[1.08, 1.50]$, $p < .001$) and questionnaire conditions ($b = 1.20$, 95\% CI $[0.98, 1.42]$, $p < .001$), whereas questionnaire and control did not differ ($b = 0.09$, $p = .330$; G1: $M = 0.75$; G2: $M = 0.85$; G3: $M = 2.04$; Appendix Fig.~\ref{fig:specificity}).

\paragraph{Exploratory prediction and mediation including goal specificity.}
Goal specificity showed a modest bivariate association with goal progress ($r = .14$, $p = .019$). However, in a participant-level extended regression including accountability, self-concordance, and condition dummies, specificity was not a significant independent predictor of goal progress ($b = 0.16$, $SE = 0.10$, $p = .105$; $n = 292$); the same pattern held in a clustered goal-level robustness model ($b = 0.09$, $p = .148$), whereas accountability remained significant ($b = 0.19$, $p = .010$). In an exploratory three-mediator AI-versus-questionnaire model adding goal specificity, indirect effects were significant for accountability ($ab = 0.14$, 95\% CI $[0.03, 0.29]$) and goal specificity ($ab = 0.27$, 95\% CI $[0.02, 0.52]$), with self-concordance remaining negligible ($n = 211$; Appendix Fig.~\ref{fig:mediation3}). The corresponding exploratory AI-versus-control model showed no individual indirect effect.

\paragraph{Exploratory shared-feature mediation via structured reflection.}
Because the AI and Questionnaire conditions were similarly high on perceived structured reflection, we ran exploratory single-mediator models using the structured-reflection manipulation check as a shared-feature mediator. Perceived structured reflection significantly mediated both the Questionnaire-versus-Control contrast ($ab = 0.28$, 95\% CI $[0.03, 0.54]$) and the AI-versus-Control contrast ($ab = 0.46$, 95\% CI $[0.15, 0.79]$), but not the AI-versus-Questionnaire contrast ($ab = 0.06$, 95\% CI $[-0.01, 0.16]$). This suggests that part of the benefit of both active interventions over no support may reflect structured reflection as a shared active ingredient, whereas the AI's added value over written reflection lies more in accountability.

\paragraph{Other exploratory outcomes.}
Exploratory secondary outcomes indicated substantially higher user satisfaction (NPS) for the AI chatbot at T1 than both the questionnaire ($d = 0.52$, $p < .001$) and the control ($d = 0.86$, $p < .001$). At T2, only the AI-versus-Control NPS contrast remained significant ($d = 0.32$, $p = .019$). By contrast, enthusiasm, goal commitment, and perceived goal difficulty did not differ significantly across conditions at T1 or T2 (all omnibus $p > .06$; Appendix A3).


\section{Discussion}\label{sec:discussion}

We conducted a preregistered three-arm RCT to test whether an AI-assisted, career-oriented goal-setting intervention improves short-term goal progress and, if so, through which psychological pathway. The AI chatbot produced significantly higher goal progress at two weeks than the no-support control ($d = 0.33$, $p = .016$; cf.\ meta-analytic $d = 0.34$ for goal-setting interventions on behaviour change~\cite{epton2017unique}), and this advantage was robust to the inclusion of demographic and AI-familiarity covariates. Compared with the written questionnaire, the AI did not significantly improve overall goal progress, but it increased perceived accountability. Exploratory analyses further suggest that part of the advantage of both active interventions over no support may reflect structured reflection as a shared active ingredient. The clearest evidence for the AI's added value over written reflection therefore points to social accountability: accountability mediated the AI-versus-questionnaire contrast, whereas self-concordance was not supported. Exploratory analyses further suggest that the AI helped participants formulate more specific goals and surfaced broader goal content, but these findings should be interpreted more cautiously.

\paragraph{Social accountability as the mechanism for conversational AI's advantage.}
The strongest finding is the accountability effect. The AI chatbot generated substantially higher perceived accountability than both the questionnaire ($d = 0.43$) and the control ($d = 0.68$), despite the questionnaire covering similar reflective topics. In the AI-versus-questionnaire contrast, accountability mediated the treatment effect (indirect effect $= 0.15$, 95\% CI $[0.04, 0.31]$), whereas self-concordance did not. This pattern suggests that the AI's clearest added value over structured written reflection lies in making users feel answerable for subsequent progress. These results extend the Computers Are Social Actors framework~\cite{nass1994computers} to goal pursuit: a chatbot that asks follow-up questions, reacts contingently, and invites future check-ins may function as a perceived social agent~\cite{epley2007seeing} to which participants feel accountable, triggering dynamics usually associated with human coaches or accountability partners~\cite{harkin2016does}.

The contrast pattern is also informative. The accountability pathway was clearest for AI versus Questionnaire, whereas the AI-versus-Control contrast showed a significant total effect without a significant individual indirect effect. This suggests that the AI's advantage over no support may reflect a broader bundle of features than the measured T1 mediators captured, whereas the questionnaire comparison more cleanly tests the conversational added value of the intervention.

\paragraph{Explaining the self-concordance null.}
The absence of self-concordance effects across conditions remains informative. The null is unlikely to reflect a simple floor effect: scores ranged widely, yet neither the AI nor the questionnaire reliably shifted self-concordance at T1 or T2. One plausible explanation is that participants in all conditions selected their own goals, leaving relatively limited room for the intervention to improve motivational alignment. Another is timescale: self-concordance may matter more for sustained effort over months than for initial progress over two weeks, where accountability-based momentum is more salient. A post hoc regression reported in Appendix A3 supported this interpretation only weakly and ambiguously: T2 self-concordance was positively associated with concurrent T2 progress, but the intervention did not differentially increase self-concordance. Exploratory subgroup analyses further suggested that within G3, participants who surfaced at least one non-career goal reported higher self-concordance than career-only G3 participants at both T1 ($d = 0.44$, $p = .033$) and T2 ($d = 0.53$, $p = .011$). We therefore view self-concordance not as the main mechanism of the AI effect in this study, but as a potentially relevant construct whose role may depend on follow-up length and on the kinds of goals the conversation surfaces.

\paragraph{Goal specificity as an exploratory finding.}
AI chatbot goals were much more specific than those in the questionnaire or control conditions, replicating the quality advantage documented by Schimpf, Maier, et al.~\cite{schimpf2025goal} and extending it to an initial goal-setting context with follow-up. This suggests that conversational AI can help users translate broad intentions into more concrete targets by probing vague language and requesting clearer timeframes or deliverables. However, the mechanism evidence was mixed in the refreshed analyses: specificity showed a modest bivariate association with progress and a significant indirect path in the exploratory AI-versus-questionnaire mediation, but it was not a robust independent predictor in the extended regression models. The most defensible interpretation is therefore that conversational AI likely improves goal formulation, and that specificity is a plausible mechanism worth preregistered follow-up rather than a settled explanation of the main effect.

\paragraph{The AI goal setting process expands goal scope beyond the career domain.}
A second exploratory implication concerns what the AI goal setting process surfaced. The LLM-based goal-domain classification revealed that 40.3\% of coded goals in G3 were non-career, compared with 13.2\% in G1 and 10.6\% in G2 ($\chi^2(2) = 96.75$, $p < .001$). This pattern is consistent with the chatbot's implementation: Leon was instructed to act as a career coach and help participants set meaningful career-related goals, but also to prioritise the goals users themselves identified as most important and not force externally derived goals onto them. When participants resisted the career framing or indicated that other priorities mattered more, the system was designed to follow that lead while still pushing for goals that felt personally meaningful and genuinely owned. Descriptive sensitivity analyses did not indicate that broader goal scope was associated with lower short-term goal progress, but it does complicate a strict career-only interpretation of the intervention. Rather than treating this purely as noise, we view it as evidence that conversational AI goal setting may uncover priorities participants regard as genuinely meaningful, albeit in a way that future studies should measure and constrain more explicitly depending on the research question.

\paragraph{Theoretical and practical implications.}
These findings carry implications for both theory and practice. Theoretically, they suggest that the value of conversational AI goal setting may be primarily social rather than motivational: in this study, the strongest supported pathway was accountability, not self-concordance. This qualifies theories that would expect guided reflection to improve goal pursuit mainly by increasing motivational alignment. Practically, the likely active ingredients appear to be features that make the system feel socially present and action-oriented: a named persona, contingent follow-up questions, explicit goal articulation, and an invitation to future follow-up. More tentatively, conversational scaffolding may also help users formulate more concrete goals, but that proposition now needs prospective testing. For organisations and career services looking to scale goal-setting support, this is encouraging: a well-designed chatbot may reproduce part of coaching's accountability function at very low marginal cost (cf.\ meta-analytic $d = 0.28$--$0.59$ for chatbot behaviour-change interventions across health domains~\cite{singh2023systematic}). AI chatbot participants also reported substantially higher user satisfaction (NPS) at T1 than both comparison conditions ($d = 0.86$ vs.\ control, $p < .001$; full statistics in Appendix A3), suggesting a practically attractive profile of both engagement and short-term effectiveness.

\paragraph{Limitations and future directions.}
Several limitations should be noted. The two-week follow-up window is short; longer observation periods are needed to determine whether accountability effects persist or attenuate, and whether self-concordance differences emerge over longer periods. The Prolific sample of employed adults (predominantly UK and US) limits generalisability to students, unemployed individuals, or non-English speakers. The questionnaire condition also imposed a 20-minute minimum completion time, which likely increased friction at T1 and may have affected dropout and comparability between the two active interventions. The study used a single LLM configuration, and results may differ across models or prompting strategies. In addition, the specificity and goal-domain analyses were exploratory and relied on LLM-based coding; these findings should be replicated with preregistered analyses and, ideally, human-validation checks. Finally, although the intervention was career-oriented, the AI condition surfaced broader goal content than the static conditions. The primary outcome should therefore be interpreted as progress on participants' selected goals rather than as a purely career-only outcome.

Future research should include active human coaching comparisons, longer follow-ups, preregistered tests of goal-quality mechanisms, and dismantling studies designed to isolate which conversational features---turn-taking, social presence, explicit goal articulation, or future commitment requests---generate accountability.


\section{Methods}\label{sec:methods}

This study was approved by the Department of Engineering Ethics Board at the University of Cambridge. All participants provided informed consent prior to participation.

\subsection{Participants and design}\label{subsec:participants}

We pre-registered a three-arm between-subjects RCT and recruited participants via Prolific. Eligibility criteria were: UK or US residents, aged 18--50, currently employed (full- or part-time), and fluent English speakers. The preregistered recruitment target was $N = 402$ (approximately $n = 134$ per condition), chosen to yield a retained analytic sample sufficient for the parallel mediation analyses (Monte Carlo power analysis: conservative assumptions path~$a$: $r = .20$, path~$b$: $r = .25$, $\alpha = .05$; power $> .90$ at $N \approx 351$). Final data collection yielded $N = 517$ randomised participants. The study was preregistered at \url{https://archive.org/details/osf-registrations-6awmq-v1} prior to data collection. Participants were randomly assigned to one of the three conditions upon entering the study platform.

\subsection{Measures}\label{subsec:measures}

\paragraph{Primary and preregistered measures.} Goal progress was assessed at T2 using three items adapted from Saunders et al.~\cite{saunders2022longitudinal}: ``I have made a lot of progress toward my goal,'' ``I feel like I am on track with my goal plan,'' and ``I feel like I have achieved my goal'' (7-point scale; 1 = \emph{strongly disagree} to 7 = \emph{strongly agree}). Items were averaged across each participant's three recorded goals (9 items total; $\alpha = .86$). Self-concordance was measured following Sheldon and Kasser~\cite{sheldon1995self,sheldon1999role}. Participants rated each goal on four motivational dimensions---external, introjected, identified, and intrinsic regulation---using 5-point scales (1 = \emph{disagree} to 5 = \emph{agree}). Self-concordance was computed as (intrinsic + identified) $-$ (introjected + external) for each goal, then averaged across goals; the scale was re-administered at T2, but T1 self-concordance was the preregistered mediator. Perceived accountability was assessed with four items adapted from Royle et al.~\cite{royle2005interactive} on a 7-point scale (1 = \emph{strongly disagree} to 7 = \emph{strongly agree}; e.g., ``I felt that I would be held accountable for achieving my goal by the goal-setting system''; $\alpha = .83$). The scale was also re-administered at T2 to assess persistence of accountability effects. Two single-item manipulation checks were administered at T1 on 7-point scales: perceived interactivity (``The goal-setting process involved a two-way, interactive conversation'') and perceived structured reflection (``Before setting my goals, I was guided through a structured reflection'').

\paragraph{Additional measures and exploratory coding.} Enthusiasm was assessed at T2 using three items per goal from Warr~\cite{warr1990measurement} ($\alpha = .89$). Goal commitment was measured with one item per goal at T1 and T2~\cite{hollenbeck1989investigation}. Perceived goal difficulty was assessed with one item per goal at T1~\cite{steers1976factors}. Net Promoter Score (NPS; ``How likely are you to recommend this goal-setting system to a friend or colleague?'', 0--10 scale) was collected at T1 and T2. Goal specificity was assessed post hoc by coding each participant's recorded goal texts using an LLM-based rubric adapted from Schimpf, Maier, et al.~\cite{schimpf2025goal}. Each goal was rated on two dimensions: time frame (0 = no time frame; 1 = vague time frame; 2 = specific date or concrete period) and measurability (0 = no measurable outcome; 1 = vague outcome; 2 = concrete measurable outcome), yielding a total specificity score of 0--4 per goal, averaged across goals for participant-level analyses. Ratings were produced by a large language model (\texttt{claude-sonnet-4-6}; Anthropic) using raw goal text only, with an explicit ``[no additional context]'' placeholder and no participant- or condition-identifying information (blind to condition). One goal triggered a model refusal and was manually adjudicated using the same rubric. Goal text was absent for 31 participants who did not record written goals (G1: $n = 24$, G2: $n = 7$, G3: $n = 0$), yielding a specificity subsample of $n = 292$. Two further text-coding analyses classified each recorded goal as career-related or non-career with a broad domain label, and coded open-ended T2 feedback responses for overall helpfulness plus non-exclusive benefit themes, challenge themes, and feature requests. Full prompts are reproduced in Appendix A5, Appendix A6, and Appendix A8.

\paragraph{Covariates.} Participants reported self-rated AI knowledge~\cite{gaube2021} and generative AI use frequency~\cite{voigt2024future} at T1. Age and gender were obtained from Prolific demographic records.

\subsection{Statistical analysis}\label{subsec:stats}

All analyses were conducted in Python~3 using \texttt{scipy}, \texttt{statsmodels}, and \texttt{numpy}, and were preregistered at \url{https://archive.org/details/osf-registrations-6awmq-v1} prior to data collection. Analysis code and anonymised data will be made available at \url{https://github.com/michel-schimpf/goal-discovery-paper} upon publication.

Primary hypotheses (H1a--c) were tested with three planned pairwise contrasts on goal progress using Welch's $t$-tests (not assuming equal variances). Effects on mediators (path~$a$; H2a--b) were assessed via one-way ANOVAs. Mediator--outcome associations (path~$b$; H3a--b) were estimated by regressing goal progress simultaneously on self-concordance, accountability, and condition dummies. Mediation (H4a--c) was tested using a parallel mediation framework following Hayes~\cite{hayes2013introduction}, with 5,000 bootstrap resamples and percentile-based 95\% confidence intervals; an indirect effect was considered significant if the CI excluded zero.

As a preregistered robustness check, the primary H1 contrasts were re-estimated as an ANCOVA including age, gender, AI knowledge, and generative AI use frequency as covariates. Influence diagnostics used Cook's distance with a threshold of $4/n$; no observations were removed. Exploratory analyses included LLM-coded goal specificity, goal-domain classification, qualitative feedback coding, three-mediator mediation models adding goal specificity, and goal-level robustness models using one row per goal with standard errors clustered by participant.

\subsection{Deviations from preregistration}\label{subsec:deviations}

The final randomised sample ($N = 517$) exceeded the preregistered recruitment target ($N = 402$) and yielded unequal condition sizes, in part because attrition was especially high in the questionnaire condition due to the platform-enforced 20-minute minimum at T1. Education was preregistered as an ANCOVA covariate but was unavailable in the final dataset and therefore could not be included. No preregistered goal-progress contrasts or two-mediator mediation tests were omitted. Additional LLM-coded analyses, three-mediator models, clustered goal-level robustness models, and the post hoc regression used to probe the self-concordance null were added after preregistration and are reported as exploratory.


\backmatter

\bmhead{Appendix}
Appendix materials include: participant flow (CONSORT; A1), the exploratory three-mediator path diagram (A2), additional exploratory analyses including descriptive statistics and intercorrelations, mediation path coefficient tables, goal specificity descriptives, goal domain classification, self-concordance decomposition by goal domain, manipulation check details, and other exploratory outcomes (A3), scale reliability coefficients (A4), the large language model prompts used for goal specificity coding and goal domain classification (A5--A6), a note pointing to the archived full chatbot transcripts retained in the repository for the non-preprint version (A7), and the LLM prompt used for qualitative feedback thematic coding (A8).

\bmhead{Acknowledgements}
We thank the participants recruited via Prolific for their time.

\section*{Declarations}

\begin{itemize}
  \item \textbf{Funding:} This research was supported by the Cyber Human Lab, Department of Engineering, University of Cambridge.
  \item \textbf{Competing interests:} The authors declare no competing interests.
  \item \textbf{Ethics approval:} This study was approved by the Department of Engineering Ethics Board at the University of Cambridge.
  \item \textbf{Consent to participate:} Informed consent was obtained from all participants.
  \item \textbf{Data availability:} Anonymised data will be made available at \url{https://github.com/michel-schimpf/goal-discovery-paper} upon publication.
  \item \textbf{Code availability:} Analysis code will be made available at \url{https://github.com/michel-schimpf/goal-discovery-paper} upon publication.
\end{itemize}

\bibliography{sn-bibliography}

@article{saunders2022longitudinal,
  author  = {Saunders, Blair and Milyavskaya, Marina and Inzlicht, Michael},
  title   = {Longitudinal evidence that event related potential measures of self-regulation do not predict everyday goal pursuit},
  journal = {Nature Communications},
  year    = {2022},
  volume  = {13},
  number  = {1},
  pages   = {3201},
  doi     = {10.1038/s41467-022-30786-7}
}

@article{royle2005interactive,
  author    = {Royle, M. Todd and Hall, Angela T. and Hochwarter, Wayne A. and Perrew{\'e}, Pamela L. and Ferris, Gerald R.},
  title     = {The interactive effects of accountability and job self-efficacy on organizational citizenship behavior and political behavior},
  journal   = {Organizational Analysis},
  year      = {2005},
  volume    = {13},
  number    = {1},
  pages     = {53--72}
}

@article{warr1990measurement,
  author  = {Warr, Peter},
  title   = {The measurement of well-being and other aspects of mental health},
  journal = {Journal of Occupational Psychology},
  year    = {1990},
  volume  = {63},
  number  = {3},
  pages   = {193--210},
  doi     = {10.1111/j.2044-8325.1990.tb00521.x}
}

@article{hollenbeck1989investigation,
  author  = {Hollenbeck, John R. and Klein, Howard J. and O'Leary, Anne M. and Wright, Patrick M.},
  title   = {Investigation of the construct validity of a self-report measure of goal commitment},
  journal = {Journal of Applied Psychology},
  year    = {1989},
  volume  = {74},
  number  = {6},
  pages   = {951--956},
  doi     = {10.1037/0021-9010.74.6.951}
}

@article{steers1976factors,
  author  = {Steers, Richard M.},
  title   = {Factors affecting job attitudes in a goal-setting environment},
  journal = {Academy of Management Journal},
  year    = {1976},
  volume  = {19},
  number  = {1},
  pages   = {6--16},
  doi     = {10.2307/255443}
}

@article{gaube2021,
  author  = {Gaube, Susanne and Suresh, Harini and Raue, Martina and Merritt, Alexander and Berkowitz, Seth J. and Lermer, Eva and Coughlin, Joseph F. and Guttag, John V. and Colak, Errol and Ghassemi, Marzyeh},
  title   = {Do as {AI} say: susceptibility in deployment of clinical decision-aids},
  journal = {npj Digital Medicine},
  year    = {2021},
  volume  = {4},
  pages   = {31},
  doi     = {10.1038/s41746-021-00385-9}
}

@article{voigt2024future,
  author  = {Voigt, Julian and Strauss, Karoline},
  title   = {How future work self salience shapes the effects of interacting with artificial intelligence},
  journal = {Journal of Vocational Behavior},
  year    = {2024},
  volume  = {155},
  pages   = {104054},
  doi     = {10.1016/j.jvb.2024.104054}
}

@article{harkin2016does,
  author    = {Harkin, Benjamin and Webb, Thomas L. and Chang, Betty P. I. and Prestwich, Andrew and Conner, Mark and Kellar, Ian and Benn, Yael and Sheeran, Paschal},
  title     = {Does monitoring goal progress promote goal attainment? A meta-analysis of the experimental evidence},
  journal   = {Psychological Bulletin},
  volume    = {142},
  number    = {2},
  pages     = {198--229},
  year      = {2016},
  doi       = {10.1037/bul0000025}
}

@article{terblanche2022efficacy,
  author  = {Terblanche, Nicky and Molyn, James and de Haan, Erik and Nilsson, Viktor O.},
  title   = {Comparing artificial intelligence and human coaching goal attainment efficacy},
  journal = {PLOS ONE},
  volume  = {17},
  number  = {6},
  pages   = {e0270255},
  year    = {2022},
  doi     = {10.1371/journal.pone.0270255}
}

@article{terblanche2022comparing,
  author    = {Terblanche, Nicky and Molyn, James and de Haan, Erik and Nilsson, Viktor O.},
  title     = {Coaching at scale: investigating the efficacy of artificial intelligence coaching},
  journal   = {International Journal of Evidence Based Coaching and Mentoring},
  volume    = {20},
  number    = {2},
  pages     = {20--36},
  year      = {2022},
  doi       = {10.24384/5cgf-ab69}
}

@article{fitzpatrick2017delivering,
  author    = {Fitzpatrick, Kathleen Kara and Darcy, Alison and Vierhile, Molly},
  title     = {Delivering cognitive behavior therapy to young adults with symptoms of depression and anxiety using a fully automated conversational agent ({Woebot}): a randomized controlled trial},
  journal   = {JMIR Mental Health},
  volume    = {4},
  number    = {2},
  pages     = {e19},
  year      = {2017},
  doi       = {10.2196/mental.7785}
}

@article{aggarwal2023artificial,
  author    = {Aggarwal, Abhishek and Tam, Cheuk Chi and Wu, Dezhi and Li, Xiaoming and Qiao, Shan},
  title     = {Artificial intelligence--based chatbots for promoting health behavioral changes: systematic review},
  journal   = {Journal of Medical Internet Research},
  volume    = {25},
  pages     = {e40789},
  year      = {2023},
  doi       = {10.2196/40789}
}

@article{kuhail2023interacting,
  author    = {Kuhail, Mohammad Amin and Alturki, Nouf and Alramlawi, Salma and Alhejori, Khadijah},
  title     = {Interacting with educational chatbots: a systematic review},
  journal   = {Education and Information Technologies},
  volume    = {28},
  pages     = {973--1018},
  year      = {2023},
  doi       = {10.1007/s10639-022-11177-3}
}

@article{schimpf2025goal,
  author    = {Schimpf, Michel and Maier, Sebastian and Wyrowski, Anton and
               Christoforakos, Lara and Feuerriegel, Stefan and Bohn{\'e}, Thomas},
  title     = {Supporting Effective Goal Setting with {LLM}-Based Chatbots},
  journal   = {arXiv preprint arXiv:2602.08636},
  year      = {2026},
  doi       = {10.48550/arXiv.2602.08636},
  eprint    = {2602.08636},
  archivePrefix = {arXiv},
  primaryClass  = {cs.HC},
  url       = {https://arxiv.org/abs/2602.08636}
}

@article{sheldon1995self,
  author    = {Sheldon, Kennon M. and Kasser, Tim},
  title     = {Coherence and congruence: two aspects of personality integration},
  journal   = {Journal of Personality and Social Psychology},
  volume    = {68},
  number    = {3},
  pages     = {531--543},
  year      = {1995},
  doi       = {10.1037/0022-3514.68.3.531}
}

@article{sheldon1999role,
  author    = {Sheldon, Kennon M. and Elliot, Andrew J.},
  title     = {Goal striving, need satisfaction, and longitudinal well-being: the self-concordance model},
  journal   = {Journal of Personality and Social Psychology},
  volume    = {76},
  number    = {3},
  pages     = {482--497},
  year      = {1999},
  doi       = {10.1037/0022-3514.76.3.482}
}

@incollection{frink1998toward,
  author    = {Frink, Dwight D. and Klimoski, Richard J.},
  title     = {Toward a theory of accountability in organizations and human resource management},
  booktitle = {Research in Personnel and Human Resources Management},
  editor    = {Ferris, Gerald R.},
  volume    = {16},
  pages     = {1--51},
  publisher = {JAI Press},
  address   = {Greenwich, CT},
  year      = {1998}
}

@article{nass1994computers,
  author    = {Nass, Clifford and Steuer, Jonathan and Tauber, Ellen R.},
  title     = {Computers are social actors},
  journal   = {Proceedings of the {SIGCHI} Conference on Human Factors in Computing Systems},
  pages     = {72--78},
  year      = {1994},
  doi       = {10.1145/191666.191703}
}

@book{locke1990theory,
  author    = {Locke, Edwin A. and Latham, Gary P.},
  title     = {A Theory of Goal Setting and Task Performance},
  publisher = {Prentice Hall},
  address   = {Englewood Cliffs, NJ},
  isbn      = {0139131388},
  year      = {1990}
}

@article{lerner1999accounting,
  author  = {Lerner, Jennifer S. and Tetlock, Philip E.},
  title   = {Accounting for the effects of accountability},
  journal = {Psychological Bulletin},
  volume  = {125},
  number  = {2},
  pages   = {255--275},
  year    = {1999},
  doi     = {10.1037/0033-2909.125.2.255}
}

@article{lockelatham2002,
  author  = {Locke, Edwin A. and Latham, Gary P.},
  title   = {Building a practically useful theory of goal setting and task motivation: A 35-year odyssey},
  journal = {American Psychologist},
  volume  = {57},
  number  = {9},
  pages   = {705--717},
  year    = {2002},
  doi     = {10.1037/0003-066X.57.9.705}
}

@article{dehaan2023,
  author  = {de Haan, Erik and Nilsson, Viktor O.},
  title   = {What can we know about the effectiveness of coaching? {A} meta-analysis based only on randomized controlled trials},
  journal = {Academy of Management Learning \& Education},
  volume  = {22},
  number  = {4},
  pages   = {641--661},
  year    = {2023},
  doi     = {10.5465/amle.2022.0107}
}

@article{deciryan2000what,
  author  = {Deci, Edward L. and Ryan, Richard M.},
  title   = {The ``what'' and ``why'' of goal pursuits: Human needs and the self-determination of behavior},
  journal = {Psychological Inquiry},
  volume  = {11},
  number  = {4},
  pages   = {227--268},
  year    = {2000},
  doi     = {10.1207/S15327965PLI1104_01}
}

@article{salvi2025,
  author    = {Salvi, Francesco and {Horta Ribeiro}, Manoel and Gallotti, Riccardo and West, Robert},
  title     = {On the conversational persuasiveness of {GPT}-4},
  journal   = {Nature Human Behaviour},
  volume    = {9},
  pages     = {1645--1653},
  year      = {2025},
  doi       = {10.1038/s41562-025-02194-6}
}

@article{mohr2011supportive,
  author  = {Mohr, David C. and Cuijpers, Pim and Lehman, Kenneth},
  title   = {Supportive accountability: {A} model for providing human support
             to enhance adherence to {eHealth} interventions},
  journal = {Journal of Medical Internet Research},
  volume  = {13},
  number  = {1},
  pages   = {e30},
  year    = {2011},
  doi     = {10.2196/jmir.1602}
}

@article{nass2000machines,
  author  = {Nass, Clifford and Moon, Youngme},
  title   = {Machines and mindlessness: Social responses to computers},
  journal = {Journal of Social Issues},
  volume  = {56},
  number  = {1},
  pages   = {81--103},
  year    = {2000},
  doi     = {10.1111/0022-4537.00153}
}

@article{epton2017unique,
  author  = {Epton, Tracy and Currie, Siobh{\'a}n and Armitage, Christopher J.},
  title   = {Unique effects of setting goals on behavior change: Systematic review
             and meta-analysis},
  journal = {Journal of Consulting and Clinical Psychology},
  volume  = {85},
  number  = {12},
  pages   = {1182--1198},
  year    = {2017},
  doi     = {10.1037/ccp0000260}
}

@article{demszky2023using,
  author  = {Demszky, Dorottya and Yang, Diyi and Yeager, David S. and Bryan,
             Christopher J. and Clapper, Margarett and Chandhok, Susannah and
             Eichstaedt, Johannes C. and Hecht, Cameron and Jamieson, Jeremy and
             Johnson, Meghann and Jones, Michaela and {Krettek-Cobb}, Danielle and
             Lai, Leslie and {JonesMitchell}, Nirel and Ong, Desmond C. and Dweck, Carol S. and
             Gross, James J. and Pennebaker, James W.},
  title   = {Using large language models in psychology},
  journal = {Nature Reviews Psychology},
  volume  = {2},
  number  = {11},
  pages   = {688--701},
  year    = {2023},
  doi     = {10.1038/s44159-023-00241-5}
}

@article{singh2023systematic,
  author  = {Singh, Ben and Olds, Timothy and Brinsley, Jacinta and Dumuid,
             Dot and Virgara, Rosa and Matricciani, Lisa and Watson, Amanda
             and Szeto, Kimberley and Eglitis, Emily and Miatke, Aaron and
             Simpson, Catherine E.~M. and Vandelanotte, Corneel and Maher, Carol},
  title   = {Systematic review and meta-analysis of the effectiveness of
             chatbots on lifestyle behaviours},
  journal = {npj Digital Medicine},
  volume  = {6},
  pages   = {118},
  year    = {2023},
  doi     = {10.1038/s41746-023-00856-1}
}

@article{epley2007seeing,
  author  = {Epley, Nicholas and Waytz, Adam and Cacioppo, John T.},
  title   = {On seeing human: {A} three-factor theory of anthropomorphism},
  journal = {Psychological Review},
  volume  = {114},
  number  = {4},
  pages   = {864--886},
  year    = {2007},
  doi     = {10.1037/0033-295X.114.4.864}
}

@book{hayes2013introduction,
  author    = {Hayes, Andrew F.},
  title     = {Introduction to Mediation, Moderation, and Conditional Process Analysis: {A} Regression-Based Approach},
  publisher = {Guilford Publications},
  address   = {New York, NY},
  isbn      = {9781609182304},
  year      = {2013}
}


\clearpage
\setcounter{section}{0}
\renewcommand{\thesection}{A\arabic{section}}
\renewcommand{\theHsection}{A\arabic{section}}
\setcounter{figure}{0}
\renewcommand{\thefigure}{A\arabic{figure}}
\renewcommand{\theHfigure}{A\arabic{figure}}
\setcounter{table}{0}
\renewcommand{\thetable}{A\arabic{table}}
\renewcommand{\theHtable}{A\arabic{table}}

\begin{center}
  {\Large\bfseries Appendix}\\[4pt]
  {\normalsize AI-Assisted Goal Setting Improves Goal Progress Through\\
  Social Accountability}
\end{center}

\vspace{1em}

\noindent\textbf{Contents:} A1~Participant flow (CONSORT). A2~Exploratory three-mediator model. A3~Additional exploratory analyses. A4~Scale reliability. A5~Goal specificity coding prompt. A6~Goal domain classification prompt. A7~Archived full chatbot transcripts (repository file). A8~Qualitative feedback coding prompt.

\bigskip

\begin{figure}[h]
  \centering
  \includegraphics[width=\textwidth]{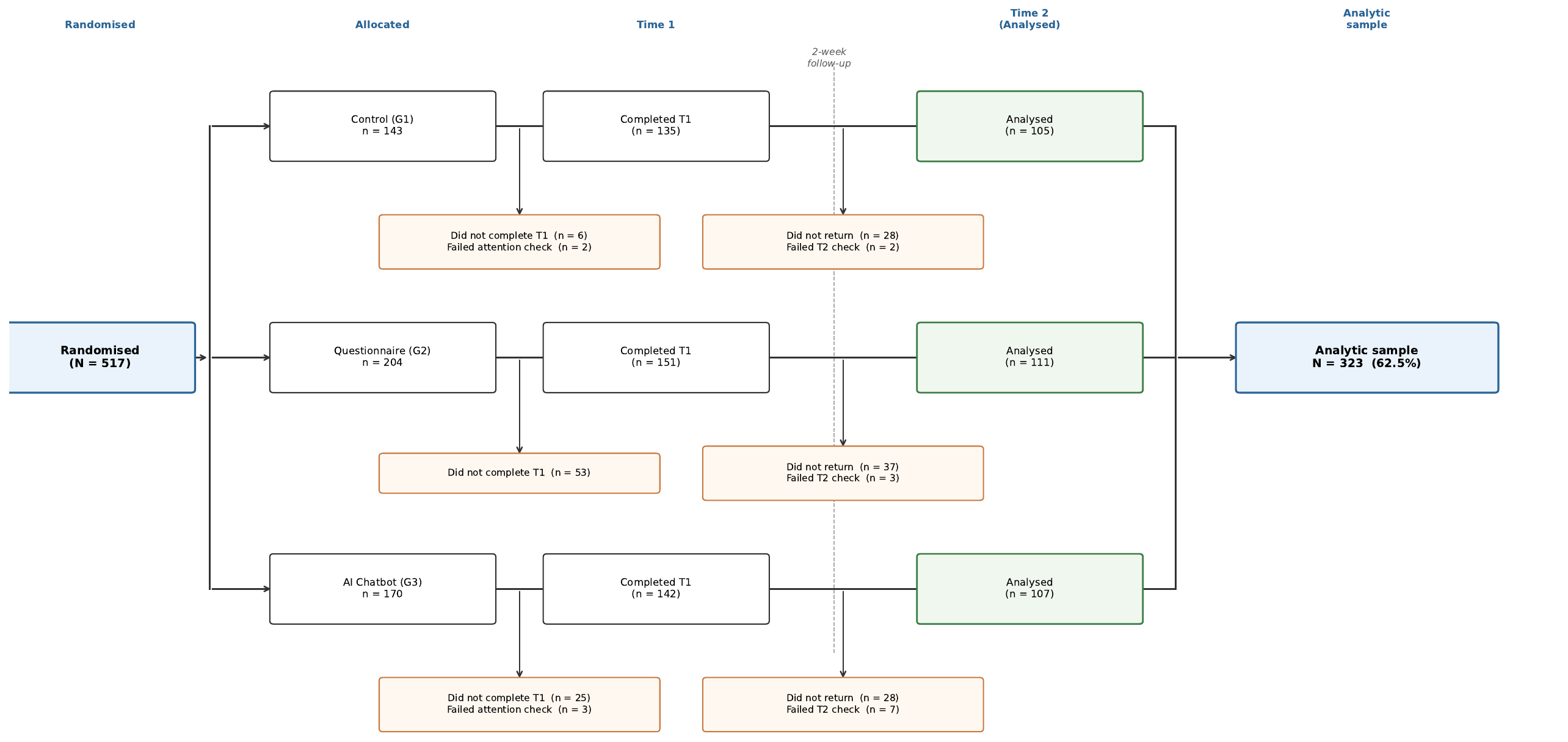}
  \caption{Participant flow. $N = 517$ randomised; 89 excluded at Time~1 (84 did not complete the session, 5 failed the attention check), leaving $n = 428$ who completed Time~1. The high T1 dropout in the Questionnaire condition ($n = 53$, all non-completions) reflects the platform-enforced 20-minute minimum engagement period. At Time~2, 93 did not return and 12 failed the attention check, yielding an analytic sample of $n = 323$ (62.5\% of randomised).}
  \label{fig:consort}
\end{figure}

\clearpage

\begin{figure}[h]
  \centering
  \includegraphics[width=0.82\textwidth]{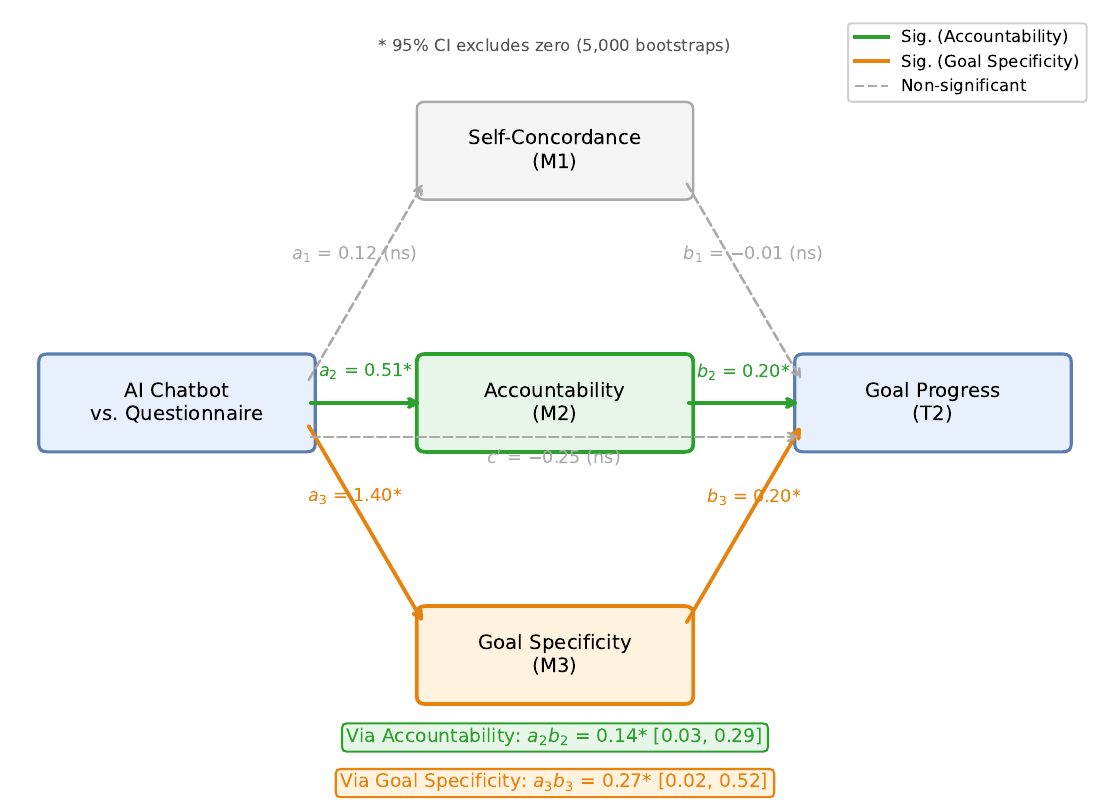}
  \caption{Exploratory parallel mediation model for the AI Chatbot versus Questionnaire contrast. Green paths: accountability; orange paths: goal specificity; grey dashed: self-concordance. Indirect effects: accountability $ab = 0.14$, 95\% CI $[0.03, 0.29]$; goal specificity $ab = 0.27$, 95\% CI $[0.02, 0.52]$. All based on 5,000 bootstraps, $n = 211$.}
  \label{fig:mediation3}
\end{figure}

\clearpage


\noindent\textbf{A3. Additional Exploratory Analyses}

\paragraph{Descriptive statistics and focal intercorrelations.}
Tables~\ref{tab:desc_stats_app} and~\ref{tab:focal_corr} report descriptive statistics for all study measures and pairwise Pearson correlations among the focal outcome and mechanism variables (analytic sample, $N = 323$; goal specificity $n = 292$).

\begin{table}[h]
\caption{Descriptive statistics for all study measures. $N = 323$ except goal specificity ($n = 292$). Self-concordance is the standard composite of intrinsic $+$ identified $-$ introjected $-$ external motivation (range $-8$ to $+8$). NPS = Net Promoter Score (0--10). Goal specificity = exploratory LLM-coded time-frame $+$ measurability (0--4).}
\label{tab:desc_stats_app}
\begin{tabular}{@{}lrrr@{}}
\toprule
Variable & $n$ & $M$ & $SD$ \\
\midrule
Goal Progress (T2)         & 323 & 3.27 & 1.33 \\
Accountability (T1)        & 323 & 3.51 & 1.27 \\
Accountability (T2)        & 323 & 3.13 & 1.51 \\
Self-Concordance (T1)      & 323 & 1.91 & 2.45 \\
Self-Concordance (T2)      & 323 & 2.05 & 3.56 \\
Goal Specificity           & 292 & 1.26 & 0.96 \\
NPS (T1)                   & 323 & 5.68 & 2.65 \\
NPS (T2)                   & 323 & 5.94 & 2.55 \\
Goal Commitment (T1)       & 323 & 3.33 & 0.58 \\
Goal Commitment (T2)       & 323 & 4.62 & 0.99 \\
Goal Difficulty (T1)       & 323 & 2.93 & 0.72 \\
Enthusiasm (T2)            & 323 & 2.56 & 0.83 \\
AI Knowledge               & 323 & 3.67 & 0.60 \\
Generative AI Use Frequency & 323 & 3.98 & 1.71 \\
\bottomrule
\end{tabular}
\end{table}

\begin{table}[h]
\caption{Intercorrelations among focal outcome and mechanism variables. $N = 323$ except goal specificity ($n = 292$). $^{*}p < .05$.}
\label{tab:focal_corr}
\begin{tabular}{@{}lrrrrrr@{}}
\toprule
Variable & 1 & 2 & 3 & 4 & 5 & 6 \\
\midrule
1.\ Goal Progress (T2) \\
2.\ Accountability (T1)   & $.19^{*}$ \\
3.\ Accountability (T2)   & $.29^{*}$ & $.50^{*}$ \\
4.\ Self-Concordance (T1) & $-.07$ & $-.01$ & $-.10$ \\
5.\ Self-Concordance (T2) & $.03$ & $-.02$ & $-.15^{*}$ & $.74^{*}$ \\
6.\ Goal Specificity      & $.14^{*}$ & $.09$ & $.13^{*}$ & $-.04$ & $-.06$ \\
\bottomrule
\end{tabular}
\vspace{2pt}
{\footnotesize Variable key: 1 = Goal Progress (T2), 2 = Accountability (T1), 3 = Accountability (T2), 4 = Self-Concordance (T1), 5 = Self-Concordance (T2), 6 = Goal Specificity.}
\end{table}

\paragraph{Manipulation checks and the H4a mediation pattern.}
The manipulation checks showed that the AI and Questionnaire conditions were similarly high on perceived structured reflection ($M = 4.45$ vs.\ $M = 4.13$), while the AI was much higher on perceived interactivity than either static condition. This supports interpreting the AI-versus-Questionnaire contrast as differing primarily in conversational interactivity, though not as a perfectly single-ingredient manipulation. Consistent with the main Results, accountability mediated this contrast in the preregistered model. By contrast, the AI-versus-Control comparison showed a significant total effect without a significant individual indirect effect, suggesting that the AI's advantage over no support reflects a broader bundle of intervention features than the measured T1 mediators captured.

\paragraph{Preregistered two-mediator model: full path coefficients (H4c).}
Table~\ref{tab:mediation_paths} reports all six ordinary least squares path coefficients for the preregistered parallel mediation model (H4c: AI Chatbot vs.\ Questionnaire, $n = 218$), equivalent to Hayes PROCESS Model~4.

\begin{table}[h]
\caption{Path coefficients for the preregistered two-mediator parallel mediation model (H4c: AI Chatbot vs.\ Questionnaire; $n = 218$). X = AI Chatbot condition dummy (1 = AI, 0 = Questionnaire). Mediators: M$_1$ = Self-Concordance (T1); M$_2$ = Accountability (T1). Outcome: Goal Progress (T2). All paths estimated via ordinary least squares. Indirect effects from 5,000 bootstrapped samples (seed = 42). $^{*}p < .05$.}\label{tab:mediation_paths}
\begin{tabular}{@{}llrrrr@{}}
\toprule
 & & $b$ & $SE$ & $t$ & $p$ \\
\midrule
\multicolumn{6}{@{}l}{\textit{Panel A: Paths from condition (X) to mediators}} \\
$a_1$ & X $\to$ Self-Concordance   & $0.122$ & $0.339$ & $0.36$ & $.720$ \\
$a_2$ & X $\to$ Accountability     & $0.509^{*}$ & $0.161$ & $3.17$ & $.002$ \\
\addlinespace
\multicolumn{6}{@{}l}{\textit{Panel B: Paths from mediators to outcome (controlling for X)}} \\
$b_1$ & Self-Concordance $\to$ Goal Progress & $-0.016$ & $0.036$ & $-0.44$ & $.659$ \\
$b_2$ & Accountability $\to$ Goal Progress   & $0.290^{*}$ & $0.076$ & $3.81$ & $< .001$ \\
\addlinespace
\multicolumn{6}{@{}l}{\textit{Panel C: Condition effects on Goal Progress}} \\
$c$ & Total effect    & $0.113$ & $0.185$ & $0.61$ & $.540$ \\
$c'$ & Direct effect  & $-0.032$ & $0.184$ & $-0.18$ & $.860$ \\
\midrule
\multicolumn{6}{@{}l}{\textit{Panel D: Bootstrapped indirect effects}} \\
$a_1 b_1$ & Via Self-Concordance & \multicolumn{4}{l}{$ab = -0.002$, 95\% CI $[-0.039,\ 0.027]$, ns} \\
$a_2 b_2$ & Via Accountability   & \multicolumn{4}{l}{$ab = 0.148^{*}$, 95\% CI $[0.036,\ 0.309]$} \\
          & Total indirect       & \multicolumn{4}{l}{$ab = 0.146^{*}$, 95\% CI $[0.027,\ 0.313]$} \\
\bottomrule
\end{tabular}
\end{table}

\begin{figure}[h]
  \centering
  \includegraphics[width=0.92\textwidth]{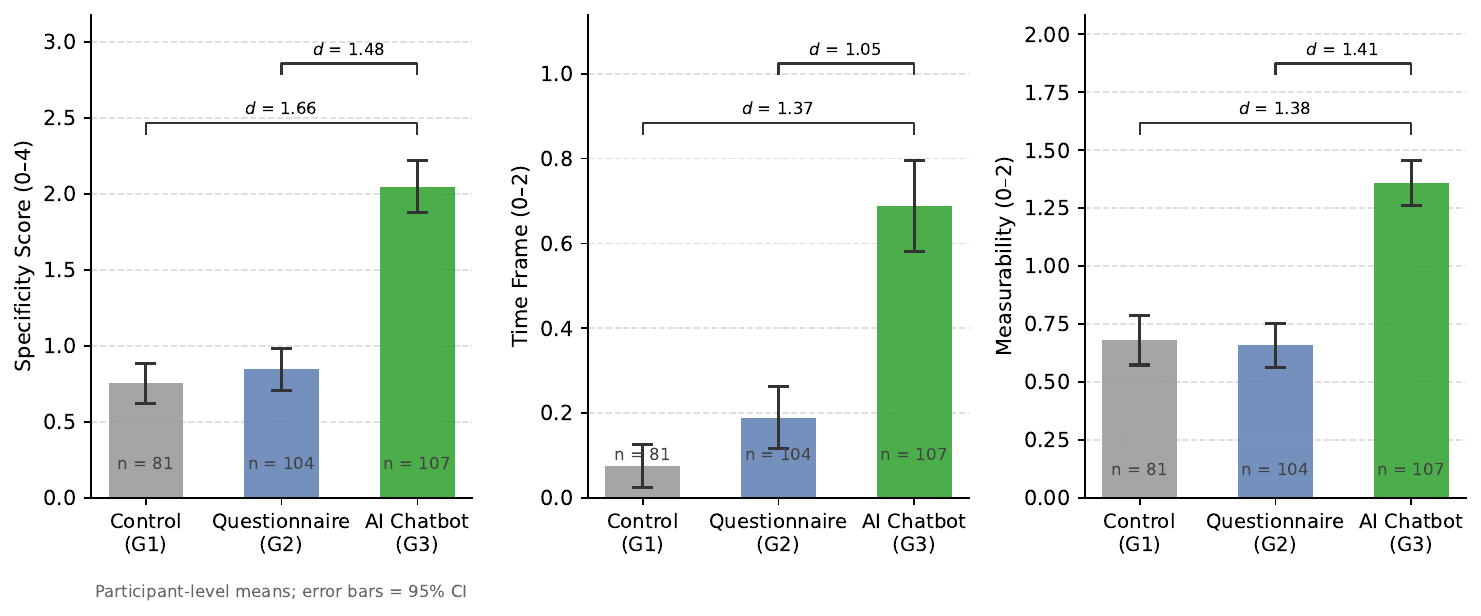}
  \caption{Descriptive goal specificity by condition (exploratory LLM coding). Left: total specificity score (0--4); centre: time-frame subscale (0--2); right: measurability subscale (0--2). Bars show participant-level means averaged across each participant's coded goals; error bars are 95\% confidence intervals. Pairwise Cohen's $d$ values for AI versus Control and AI versus Questionnaire are annotated descriptively.}
  \label{fig:specificity}
\end{figure}

\paragraph{Descriptive goal specificity plots.}
Figure~\ref{fig:specificity} visualises the condition differences in exploratory goal specificity using participant-level means averaged across each participant's coded goals. These descriptive plots complement the clustered goal-level condition model reported in the main text.

\paragraph{Goal domain classification.}
We examined the proportion of non-career goals across conditions using the same LLM coding approach as goal specificity (\texttt{claude-sonnet-4-6}; Anthropic). The system prompt provided explicit edge-case guidance (e.g., ``run a sale / run-through of interview answers'' $\to$ career; ``go running / run 5k'' $\to$ non-career; social media goals $\to$ career if professional/business-facing) and returned a binary career/non-career classification plus a domain label (career, fitness\_health, family\_relationships, personal\_development, leisure, personal\_finance, other) for each goal. The coder was blind to condition.

Of 875 classified goals, 194 (22.2\%) were coded as non-career. The proportion differed significantly by condition ($\chi^2(2) = 96.75$, $p < .001$): the AI chatbot condition produced substantially more non-career goals (40.3\%, $n = 129/320$) than either the control condition (13.2\%, $n = 32/243$; $z = 7.06$, $p < .001$, $h = 0.63$) or the questionnaire condition (10.6\%, $n = 33/312$; $z = 8.56$, $p < .001$, $h = 0.71$). The questionnaire and control conditions did not differ significantly ($z = -0.94$, $p = .346$, $h = -0.08$). At the participant level, among participants with codable goal text, 65.4\% of G3 participants ($n = 70/107$) set at least one non-career goal, compared with 27.2\% of G1 ($n = 22/81$) and 23.1\% of G2 ($n = 24/104$).

The full goal-level domain breakdown across all three conditions is presented in Appendix Table~\ref{tab:domain_class}. In G3, fitness and health (48, 37.2\% of G3 non-career goals), personal finance (28, 21.7\%), and leisure (17, 13.2\%) were the most common non-career domains. In G1 and G2, non-career goals were concentrated more in personal development and miscellaneous categories. This pattern is consistent with the AI's open-ended conversational design, which prompted participants to reflect on what mattered most to them before narrowing to goals.

Descriptive sensitivity analyses further indicated that non-career goal setting did not compromise short-term goal progress. Among G3 participants, those with at least one non-career goal ($n = 70$; $M = 3.42$, $SD = 1.34$) did not differ significantly in T2 goal progress from those who set only career-coded goals ($n = 37$; $M = 3.51$, $SD = 1.40$; $t(105) = -0.32$, $p = .748$, $d = -0.07$). This does not affect the primary outcome analyses, which assess progress toward participants' selected goals irrespective of domain.

\begin{table}[h]
\caption{Goal domain classification by condition (LLM-coded, \texttt{claude-sonnet-4-6}, blind to condition). Each participant contributed up to three goals. Percentages are of total classified goals per condition. The non-career subtotal row corresponds to the aggregate used in the main text ($\chi^2(2) = 96.75$, $p < .001$; G3 vs.\ G1: $h = 0.63$; G3 vs.\ G2: $h = 0.71$).}
\label{tab:domain_class}
\begin{tabular}{@{}lccc@{}}
\toprule
Domain & G1 Control & G2 Questionnaire & G3 AI \\
\midrule
Career                           & 211 (86.8\%) & 279 (89.4\%) & 191 (59.7\%) \\
\midrule
\textit{Non-career (total)}      & \textit{32 (13.2\%)} & \textit{33 (10.6\%)} & \textit{129 (40.3\%)} \\
\quad Fitness \& health          &   3 (1.2\%)  &   3 (1.0\%)  &  48 (15.0\%)  \\
\quad Personal finance           &   5 (2.1\%)  &   4 (1.3\%)  &  28 (8.8\%)   \\
\quad Personal development       &   8 (3.3\%)  &  15 (4.8\%)  &  15 (4.7\%)   \\
\quad Leisure                    &   4 (1.6\%)  &   3 (1.0\%)  &  17 (5.3\%)   \\
\quad Family \& relationships    &   0 ---      &   3 (1.0\%)  &   9 (2.8\%)   \\
\quad Other                      &  12 (4.9\%)  &   5 (1.6\%)  &  12 (3.8\%)   \\
\midrule
Total goals                      & 243          & 312          & 320          \\
\bottomrule
\end{tabular}
\end{table}

\paragraph{Qualitative feedback thematic coding.}
At the two-week follow-up, all participants provided open-ended written feedback on how helpful they found the goal-setting system (\textit{feedback\_helpful\_t2}; 104--111 scorable responses per condition after excluding blank or trivially non-informative replies). Responses were coded using the same LLM-based approach as the goal domain analysis (\texttt{claude-sonnet-4-6}, blind to condition). Each response was assigned an overall helpfulness rating (positive, neutral, or negative), non-exclusive benefit themes, non-exclusive challenge themes, and non-exclusive feature-request themes. The full coding prompt is reproduced in Appendix A8.

Table~\ref{tab:feedback_summary} summarises the revised feedback coding outputs. Overall helpfulness ratings still did not differ significantly by condition ($\chi^2(4) = 4.95$, $p = .293$), though G3 again showed the highest proportion of positive ratings (61.0\%) and the lowest proportion of negative ratings (14.3\%). Across all three conditions, the most common reported benefit was greater clarity and focus (G1: 45.2\%; G2: 55.9\%; G3: 59.0\%; $p = .109$). Benefit-theme prevalence differed significantly only for \textit{motivation} ($\chi^2(2) = 6.02$, $p = .049$), which was mentioned more often in the two active conditions (G2: 18.0\%; G3: 18.1\%) than in the control (7.7\%). The most common challenge was \textit{insufficient support}---typically requests for more reminders, check-ins, or sustained follow-up---and this differed by condition ($\chi^2(2) = 8.14$, $p = .017$), appearing most often in G1 (42.3\%) and least often in G3 (23.8\%). Requests for reminders or check-ins were common across all conditions (G1: 32.7\%; G2: 28.8\%; G3: 19.0\%; $p = .071$), while requests for more action steps or tracking were highest in G1 and lowest in G3 ($\chi^2(2) = 8.45$, $p = .015$). Accountability was sometimes mentioned as a benefit, but did not differ significantly across conditions (G1: 14.4\%; G2: 18.0\%; G3: 13.3\%), again suggesting that open-ended feedback only partially mirrors the scale-based accountability effect.

\begin{table}[h]
\caption{Qualitative feedback coding summary by condition (exploratory; \texttt{claude-sonnet-4-6}, blind to condition). Values are percentages of scorable responses within each condition. Benefit, challenge, and request rows are non-exclusive.}
\label{tab:feedback_summary}
\begin{tabular}{@{}lccc@{}}
\toprule
Code & G1 Control & G2 Questionnaire & G3 AI \\
\midrule
Positive helpfulness & 49.0\% & 52.3\% & 61.0\% \\
Neutral helpfulness  & 26.0\% & 28.8\% & 24.8\% \\
Negative helpfulness & 25.0\% & 18.9\% & 14.3\% \\
\addlinespace
Clarity / focus                 & 45.2\% & 55.9\% & 59.0\% \\
Reflection                      & 25.0\% & 24.3\% & 21.9\% \\
Motivation                      & 7.7\%  & 18.0\% & 18.1\% \\
Accountability                  & 14.4\% & 18.0\% & 13.3\% \\
\addlinespace
Insufficient support            & 42.3\% & 35.1\% & 23.8\% \\
Forgot / low salience           & 28.8\% & 19.8\% & 19.0\% \\
Too pushy / unrealistic         & 2.9\%  & 3.6\%  & 6.7\%  \\
\addlinespace
Requested reminders / check-ins & 32.7\% & 28.8\% & 19.0\% \\
Requested action steps / tracking & 17.3\% & 10.8\% & 4.8\% \\
\bottomrule
\end{tabular}
\end{table}

\paragraph{Other exploratory outcomes.}
Net Promoter Score (NPS) differed significantly across conditions at T1 ($F(2, 320) = 19.15$, $p < .001$, $\eta^2 = .107$), with the AI chatbot rated substantially higher ($M = 6.83$, $SD = 2.53$) than both the questionnaire ($M = 5.47$, $SD = 2.66$; $d = 0.52$, $p < .001$) and the control ($M = 4.73$, $SD = 2.31$; $d = 0.86$, $p < .001$); the questionnaire and control also differed ($d = 0.29$, $p = .031$). At T2, the AI condition retained a significant pairwise advantage over the control ($d = 0.32$, $p = .019$), though the omnibus test was marginal ($F(2, 320) = 2.61$, $p = .075$), suggesting some attenuation of satisfaction over time. Enthusiasm, goal commitment, and perceived goal difficulty did not differ significantly across conditions at T1 or T2 (all omnibus $p > .06$).

\paragraph{T2 accountability: persistence of the effect.}
To assess whether the accountability advantage of the AI chatbot persisted to the two-week follow-up, we re-examined accountability measured at T2. The omnibus test was not significant ($F(2, 320) = 2.15$, $p = .118$, $\eta^2 = .013$), with means of $M = 2.90$ (Control), $M = 3.18$ (Questionnaire), and $M = 3.32$ (AI). A pairwise comparison confirmed that the AI still exceeded the control at T2 ($d = 0.29$, $p = .035$), but no longer differed from the questionnaire ($d = 0.09$, $p = .492$). Paired $t$-tests within each condition showed that AI participants' accountability declined significantly from T1 to T2 ($\Delta = -0.63$, $t(106) = -4.87$, $p < .001$), whereas changes in the questionnaire and control conditions were non-significant. Thus, the accountability boost attenuates over two weeks, though the AI condition retains a numerical advantage over the control.

\paragraph{T2 self-concordance and SC decomposition by goal domain.}
Self-concordance was re-administered at T2 using the same four regulation items as at T1, but on a 7-point response scale rather than the 5-point scale used at T1. Consistent with the T1 null, T2 self-concordance did not differ across conditions ($F(2, 320) = 0.78$, $p = .459$; $M_{\text{Control}} = 1.70$, $M_{\text{Quest}} = 2.25$, $M_{\text{AI}} = 2.20$). Paired $t$-tests within each condition found no significant change from T1 to T2 (all $p > .14$), but these T1-to-T2 comparisons should be interpreted cautiously because the response scales differed across waves. In a post hoc regression predicting T2 progress from T1 self-concordance, T2 self-concordance, T1 accountability, and condition, T2 self-concordance positively predicted T2 progress ($b = 0.07$, $p = .033$), whereas T1 self-concordance was negatively associated ($b = -0.11$, $p = .013$). The G3 subgroup decomposition reported in the main Results section confirms that T2 SC mirrors the T1 pattern: G3 participants with at least one non-career goal reported higher T2 self-concordance than G3 career-only participants ($M_{\text{T2}} = 2.81$ vs.\ $1.03$, $d = 0.53$, $p = .011$). Full means are presented in Appendix Table~\ref{tab:sc_domain}.

\begin{table}[h]
\caption{Self-concordance (SC) by condition and goal domain (exploratory). Subgroups are defined by whether participants surfaced at least one non-career goal. SC range: $-8$ to $+8$; higher scores indicate stronger autonomous motivation. The within-condition non-career versus career-only difference was clearest in G3 (T1: $d = 0.44$, $p = .033$; T2: $d = 0.53$, $p = .011$), though smaller differences were also present in G1 and G2. T1 and T2 used the same SC items but different response scales (5-point at T1; 7-point at T2), so change scores should be interpreted cautiously.}
\label{tab:sc_domain}
\begin{tabular}{@{}lccc@{}}
\toprule
Group & $n$ & T1 SC $M$ (SD) & T2 SC $M$ (SD) \\
\midrule
G1 Control (overall)                    & 105 & 1.76 (2.35) & 1.70 (3.60) \\
\quad G1 --- non-career ($\geq$1)       &  22 & 1.09 (2.65) & 0.71 (4.00) \\
\quad G1 --- career-only                &  83 & 1.94 (2.24) & 1.96 (3.47) \\
\midrule
G2 Questionnaire (overall)              & 111 & 1.92 (2.60) & 2.25 (3.61) \\
\quad G2 --- non-career ($\geq$1)       &  24 & 2.79 (2.56) & 3.17 (3.75) \\
\quad G2 --- career-only                &  87 & 1.68 (2.57) & 1.99 (3.55) \\
\midrule
G3 AI (overall)                         & 107 & 2.04 (2.40) & 2.20 (3.47) \\
\quad G3 --- non-career ($\geq$1)       &  70 & 2.39 (2.39) & 2.81 (3.37) \\
\quad G3 --- career-only                &  37 & 1.37 (2.28) & 1.03 (3.40) \\
\bottomrule
\end{tabular}
\end{table}

\paragraph{Conversation length as a dose-response check (G3 only).}
Within the AI chatbot condition ($n = 107$), we examined whether conversation duration and message count predicted the proposed mechanisms and the primary outcome. Session duration (M = 21.9~min, SD = 9.7) was positively associated with perceived accountability ($r = .22$, $p = .023$), but not with goal specificity ($r = .14$, $p = .152$) or goal progress at T2 ($r = .09$, $p = .370$). Message count was positively associated with goal specificity ($r = .24$, $p = .014$), but not with accountability ($r = .02$, $p = .866$) or goal progress ($r = .01$, $p = .886$). These descriptive patterns suggest that different aspects of conversational engagement may relate to different exploratory mechanisms, but they do not establish a dose-response relationship for the primary outcome.

\paragraph{Moderation by AI familiarity.}
We explored whether the benefit of the AI chatbot was moderated by prior AI familiarity (self-rated AI knowledge and generative AI use frequency). Neither the AI knowledge interaction ($b = 0.47$, $p = .109$) nor the generative AI use interaction ($b = 0.18$, $p = .101$) significantly moderated the AI-versus-control effect on goal progress, suggesting that the observed treatment effect does not depend strongly on users' prior experience with AI tools. This is consistent with the interpretation that the social-actor response documented by CASA theory is largely automatic and not contingent on technological sophistication.

\paragraph{Demographic moderation.}
We examined whether the AI-versus-control advantage on goal progress was moderated by gender or country of residence ($n = 210$ after excluding one consent-revoked participant). Neither the gender-by-condition interaction ($b = -0.42$, $SE = 0.37$, $p = .253$) nor the country-by-condition interaction ($b = +0.49$, $SE = 0.37$, $p = .180$) was statistically significant. The main effect of AI condition remained robust in both models (both $p < .030$). Simple slopes showed numerically larger AI effects among male participants ($d = 0.54$, $p = .023$) and UK participants ($d = 0.48$, $p = .010$) relative to female ($d = 0.19$, $p = .283$) and US participants ($d = 0.11$, $p = .641$), but given non-significant interaction terms these differences should be interpreted cautiously.


\bigskip
\noindent\textbf{A4. Scale Reliability}

Internal consistency (Cronbach's $\alpha$) was satisfactory for all multi-item scales: goal progress ($\alpha = .86$, 9 items across 3 goals), accountability ($\alpha = .83$, 4 items), and enthusiasm ($\alpha = .89$, 9 items across 3 goals). Goal commitment and goal difficulty were single-item measures per goal, averaged across goals. Self-concordance was computed as the algebraic sum of four motivation subscale items and is not typically reported as a Cronbach's $\alpha$; the scale follows the standard Sheldon \& Kasser scoring procedure.


\bigskip
\noindent\textbf{A5. Goal Specificity Coding Prompt}

\bigskip
\noindent Goals were rated by a large language model (\texttt{claude-sonnet-4-6}; Anthropic) using the following system and user prompts. The main coding run used raw goal text only, passed ``[no additional context]'' as the context field, and provided no participant or condition information. Scores were averaged across goals to obtain a participant-level specificity score (0--4).

\bigskip
\noindent\textit{System prompt:}

\begin{PromptBlock}
You are a careful research assistant coding the specificity of short-term goals for an academic study.
Score each goal on two dimensions: time-frame specificity and outcome measurability.

This is a harmless post-hoc text-classification task on participant-written goal statements. Some goals may mention scientific, medical, technical, or work-related terms. You must not give advice or instructions about those topics. You are only scoring the wording of the goal text.

Use the GOAL TEXT as the primary evidence. Additional CONTEXT may be provided, but use it only to disambiguate what the goal refers to. Do NOT use context to add missing dates, frequencies, quantities, or thresholds that are absent from the goal text.

TIME FRAME:
  0 = no time anchor in the goal text (e.g. "find a new job", "learn a new data skill", "get promoted")
  1 = broad or recurring time anchor (e.g. "this month", "next month", "next 2 years", "every day", "weekly", "3x per week")
  2 = precise deadline or clearly bounded period (e.g. "by March 31", "by end of January", "within 4 weeks", "by end of Q2", "in 1 week")

MEASURABILITY:
  0 = no clear observable deliverable or success criterion (e.g. "do better at work", "be healthier")
  1 = clear but non-quantified outcome or deliverable (e.g. "find a new job", "learn a new skill", "get certified", "create a promotion plan")
  2 = quantified, countable, thresholded, or explicitly frequency-based outcome (e.g. "submit 12 applications", "earn $2,200 per month", "work out 3x per week", "complete 5 reports and get sign-off")

Important coding rules:
- "every day", "weekly", "monthly", "3x per week" count as time_frame = 1, not 0
- Binary but concrete deliverables like "find a new job" or "get certified" should usually be measurability = 1, not 0
- If a goal has a numeric target, frequency target, named deliverable count, or explicit threshold, measurability = 2
- Be conservative: do not upgrade a score unless the goal text itself supports it

Return ONLY valid JSON exactly like: {"time_frame": 0, "measurability": 1}
\end{PromptBlock}

\noindent\textit{User prompt:}

\begin{PromptBlock}
Goal text:
"{goal}"

Additional context for disambiguation:
[no additional context]

Return JSON only
\end{PromptBlock}


\bigskip
\noindent\textbf{A6. Goal Domain Classification Prompt}

\bigskip
\noindent Goals were classified as career-related or non-career by a large language model (\texttt{claude-sonnet-4-6}; Anthropic) using the following system and user prompts. The user prompt contained only the raw goal text with no participant or condition information. Each goal received a binary classification (\texttt{is\_career}: 0 or 1) and a domain label.

\bigskip
\noindent\textit{System prompt:}

\begin{PromptBlock}
You are a research assistant classifying whether a study participant's goal is career-related or not.

The study asked participants to set "career-related goals" for the next month.

CAREER goals include: finding a job, getting a promotion, developing professional skills, completing work projects, improving work performance, growing a business, professional networking, career transitions, financial goals tied to employment income, and educational goals for career advancement.

NON-CAREER goals include: gym/exercise/fitness, physical health, diet or weight loss, family time, personal relationships, hobbies for leisure (music, art, sport), mental wellbeing unrelated to work, home maintenance, and personal lifestyle routines.

IMPORTANT EDGE CASES -- read carefully:
- "Post on social media / Instagram / LinkedIn" -> CAREER if about professional networking or growing a business/brand; NON-CAREER if purely personal
- "Read books / reading" -> CAREER if for professional development (e.g. "read books about leadership"); NON-CAREER if clearly leisure (e.g. "read 4 novels")
- "Run a sale / run a project / run-through of interview answers" -> CAREER (run = execute, not exercise)
- "Go running / run 5k / run every weekend" -> NON-CAREER (fitness)
- Work-life balance goals (e.g. "leave work on time", "stop checking email after 7pm") -> CAREER (about work habits/boundaries)
- "Exercise to have more energy for work" -> NON-CAREER (fitness, even if work-motivated)
- "Spend time with family" -> NON-CAREER
- "Manage personal finances / save money" -> personal_finance; is_career = 0 unless clearly about salary negotiation

Return ONLY valid JSON with two fields:
  is_career: 1 if career-related, 0 if non-career
  domain: one of "career", "fitness_health", "family_relationships", "personal_development", "leisure", "personal_finance", "other"

Examples:
  "Find a new job by end of month" -> {"is_career": 1, "domain": "career"}
  "Go to the gym 3x per week" -> {"is_career": 0, "domain": "fitness_health"}
  "Spend more quality time with my kids" -> {"is_career": 0, "domain": "family_relationships"}
  "Read 4 novels this month" -> {"is_career": 0, "domain": "leisure"}
  "Run an Instagram stories sale with GBP150 revenue target" -> {"is_career": 1, "domain": "career"}
  "Complete interview prep and run through answers" -> {"is_career": 1, "domain": "career"}
\end{PromptBlock}

\noindent\textit{User prompt:}

\begin{PromptBlock}
Classify this goal: "{goal}"
\end{PromptBlock}


\bigskip
\noindent\textbf{A7. Archived Full Chatbot Transcripts}

\bigskip
\noindent To keep the arXiv preprint concise, the two full example G3 transcripts previously included in this appendix have been moved to the repository file \texttt{context/appendix\_chatbot\_transcripts.tex}. That archived file preserves both transcripts verbatim: (i) a career-goal session illustrating the standard four-phase conversation and accountability check-in, and (ii) a non-career goal-setting session illustrating the broader goal-scope pattern reported in the main text.


\bigskip
\noindent\textbf{A8. Qualitative Feedback Coding Prompt}

\bigskip
\noindent The following prompt was used with \texttt{claude-sonnet-4-6} (Anthropic) to code open-ended T2 feedback responses (\textit{feedback\_helpful\_t2}) into an overall helpfulness rating, benefit themes, challenge themes, and feature-request themes. Each response was submitted independently with no condition information provided to the model.

\bigskip
\noindent\textit{System prompt:}

\begin{PromptBlock}
You are a research assistant coding open-ended qualitative feedback from a study on AI-assisted career goal setting.

Participants completed a goal-setting task (either using an AI chatbot, a written questionnaire, or no support) and were asked at a two-week follow-up: "How helpful was the goal-setting system for you?"

Code each response into FOUR fields:

1. helpfulness
- "positive": overall clearly positive/helpful
- "neutral": mixed, ambivalent, or only mildly helpful
- "negative": overall clearly unhelpful/negative

2. benefit_themes
Choose ALL that apply from:
- "accountability": mentions feeling accountable, responsible, answerable, or like they made a promise
- "clarity_focus": mentions clearer goals, better focus, more realistic goals, narrowed priorities, or keeping goals in mind
- "reflection": mentions reflection, thinking about priorities, why the goals matter, or self-understanding
- "motivation": mentions greater motivation, drive, energy, or willingness to act
- "action_planning": mentions breaking goals into steps, manageable chunks, concrete plans, or knowing what to do next
- "social_interaction": mentions liking the conversation, chat, interactivity, or feeling supported by talking it through
- "autonomy_meaning": mentions surfacing what mattered most, personally meaningful goals, or goals that felt genuinely theirs

3. challenge_themes
Choose ALL that apply from:
- "forgot_or_low_salience": forgot about the goals/process, lost track, or it did not stay top-of-mind
- "no_added_value": says it was not especially helpful, not better than doing it alone, or had little added value
- "too_pushy_or_unrealistic": says it felt pushy, pressured, bothersome, or pushed unrealistic timelines/goals
- "too_generic_or_repetitive": says it felt repetitive, basic, same-y, tick-box, or too generic
- "insufficient_support": says it lacked reminders, check-ins, guidance, resources, encouragement, or follow-through support
- "time_burden": says it took too long, felt tedious, or was burdensome to complete

4. request_themes
Choose ALL that apply from:
- "reminders_or_checkins": explicitly asks for reminders, notifications, revisiting, or follow-up check-ins
- "action_steps_or_tracking": asks for smaller steps, progress tracking, tick-boxes, or task breakdowns
- "advice_or_feedback": asks for tips, tailored advice, encouragement, resources, or feedback on progress
- "more_interaction_or_personalization": asks for more conversation, richer interaction, or more personalized support

Return ONLY valid JSON exactly like:
{
  "helpfulness": "neutral",
  "benefit_themes": ["reflection"],
  "challenge_themes": ["insufficient_support"],
  "request_themes": ["reminders_or_checkins"]
}
\end{PromptBlock}

\noindent\textit{User prompt:}

\begin{PromptBlock}
Code this feedback response: "{text}"
\end{PromptBlock}

\end{document}